\documentclass{article}

\usepackage[utf8]{inputenc}
\usepackage{todonotes}
\usepackage{authblk}
\usepackage{url}
\usepackage{hyperref}
\usepackage{paralist}
\usepackage[capitalise]{cleveref}
\usepackage{tabularx}
\usepackage{booktabs}
\usepackage{caption}
\usepackage[sorting=none]{biblatex}
\usepackage{microtype}
\usepackage[frozencache=true,cachedir=minted-cache]{minted}

\bibliography{document}

\title{A Metadata-Based Ecosystem to Improve the FAIRness of Research Software}

\author[1,*]{Patrick Kuckertz}
\author[1,5]{Jan Göpfert}
\author[2]{Oliver Karras}
\author[1]{David Neuroth}
\author[1]{Julian Schönau}
\author[1]{Rodrigo Pueblas}
\author[3]{Stephan Ferenz}
\author[2]{Felix Engel}
\author[1]{Noah Pflugradt}
\author[1]{Jann M. Weinand}
\author[3]{Astrid Nieße}
\author[2,4]{Sören Auer}
\author[1,5]{Detlef Stolten}

\affil[1]{Forschungszentrum Jülich GmbH, Institute of Energy and Climate Research – Techno-economic Systems Analysis (IEK-3), 52425 Jülich, Germany}
\affil[2]{TIB - Leibniz Information Centre for Science and Technology, 30167 Hanover, Germany}
\affil[3]{Department for Computer Science, Carl von Ossietzky University of Oldenburg, 26129 Oldenburg, Germany}
\affil[4]{University of Hannover, L3S Research Center, 30167 Hannover, Germany}
\affil[5]{RWTH Aachen University, Chair for Fuel Cells, Faculty of Mechanical Engineering, 52062 Aachen, Germany}
\affil[*]{Correspondence: p.kuckertz@fz-juelich.de}

\date{2023}

\begin{document}

\maketitle

\begin{abstract}
    The reuse of research software is central to research efficiency and academic exchange. The application of software enables researchers with varied backgrounds to reproduce, validate, and expand upon study findings. Furthermore, the analysis of open source code aids in the comprehension, comparison, and integration of approaches. Often, however, no further use occurs because relevant software cannot be found or is incompatible with existing research processes. This results in repetitive software development, which impedes the advancement of individual researchers and entire research communities. In this article, the \emph{DataDesc} ecosystem is presented – an approach to describing data models of software interfaces with detailed and machine-actionable metadata. In addition to a specialized metadata schema, an exchange format and support tools for easy collection and the automated publishing of software documentation are introduced. This approach practically increases the FAIRness, i.e., findability, accessibility, interoperability, and so the reusability of research software, as well as effectively promotes its impact on research.
\end{abstract}

\textbf{Keywords:} Research Data Management (RDM), FAIR, software metadata, interface description, semantic software description, software publication, software reuse, machine-interpretable, application profile, CodeMeta

\section{Introduction}
\label{sec:Introduction}

Research in many academic disciplines relies on computational methods, to the degree that the utilization of software has become integral in numerous fields. Thus, the efficient discovery and reuse of research software is essential for academic progress and communication. Furthermore, the examination of open source code aids in the comprehension, comparison, and integration of methodologies, and the application of software enables users with various academic backgrounds to replicate, validate, and build upon study findings. Scientific software publications are also becoming increasingly important for measuring the research impact and so for the reputation of individual researchers~\cite{anzt2021crediting, smith2016software}.

Finding compatible software that meets researchers’ content requirements and integrates seamlessly into existing research workflows remains a significant challenge~\cite{kelley2021framework}. Currently available software metadata schemas, such as\emph{CodeMeta}~\cite{CodeMeta}, only focus on general information and omit detailed technical descriptions of interfaces, which are important for interoperability and subsequent use~\cite{druskat2022software}. At most, such information can be found on software documentation sites, where it is neither standardized nor machine-actionable. Furthermore, metadata is stored and exchanged in various formats, from which no standardized exchange format has yet been developed, that would allow the broad reuse of metadata once it has been captured~\cite{lamprecht_towards_2020}. Therefore, in order to make a software known on various platforms and increase its impact, metadata must often be repeatedly collected for each platform separately, which greatly increases documentation effort, which is already perceived to be high. At the same time, the broad dissemination of metadata is essential for the long-term discoverability and subsequent use of software~\cite{habermann2020metadata}. As a result, researchers must invest considerable effort in both  documenting and publishing metadata, as well as finding and integrating research software. Every time software is not found and reused and instead redundantly developed, a significant increase in avoidable programming, documentation and maintenance efforts is imposed.

To address these issues, adaptations of the FAIR Guiding Principles, which aim to increase the findability, accessibility, interoperability, and reusability of research data~\cite{wilkinsonFAIRGuidingPrinciples2016}, were recently adopted specifically for research software~\cite{barker2022introducing, katz_taking_2021}. Amongst other things, these principles require research software to be registered and indexed in searchable platforms, and annotated with rich metadata. In order to increase the interoperability of software components, the metadata must include interface definitions of modular software architectures, making interoperability the most challenging amongst the four high-level principles. On the one hand, all metadata must comply with domain-relevant community standards in order to be easily understandable for researchers. On the other, the metadata must be machine-actionable for automated software discovery. In practice, however, it is unclear how the postulated abstract principles may be put into action~\cite{hasselbring_fair_2020}.

The DataDesc ecosystem presented in this article is a practical approach to improving the interoperability and findability of research software. It centers around a software metadata schema that describes the data models on which software interfaces are based. In order to capture characteristics that are usually only described in documentation, metadata elements from various established schemas were reused, combined, and supplemented with new ones. In addition, the ecosystem provides an exchange format in which this information is mapped in a machine-actionable manner. The hierarchical data structure of the OpenAPI standard was chosen as its basis to facilitate its reuse in automated processes. Finally, it includes a toolset that makes it easy to capture and publish software metadata from the source code.

The remainder of this article is structured as follows: \Cref{sec:Related_Work} presents a review of existing software description schemas, addressing different formats in the tension between metadata and documentation. Furthermore, automated description tools and software publication platforms are compared on the basis of the metadata formats they generate or use. \Cref{sec:The_DataDesc_Ecosystem} explains the different components of the DataDesc ecosystem. First, the DataDesc schema is described along with the typical data flow between individual interface components of research software on the basis of its contents, formats, value ranges, and structures. Then, an explanation of the structure of the exchange format and the individual tools that support metadata generation is given. Finally, pipelines to publication platforms are described with which the metadata can be disseminated in a partially automated way. In \Cref{sec:Application_Case}, the presented approach is exemplarily applied to the Framework for Integrated Energy System Assessment (ETHOS.FINE)~\cite{WELDER20181130} modeling framework from the energy domain, whereupon the strengths and limitations of the DataDesc schema in particular are discussed. \Cref{sec:Summary_and_Conclusions} concludes with a summary of the key characteristics of the presented approach and provides an outlook on future work.

\section{Related Work}
\label{sec:Related_Work}

An overview of current software description schemas is provided in \Cref{subsec:Software_Description_Schemas}.
Additionally, software publishing platforms and automated description tools are contrasted in sections \ref{subsec:Software_Description_Tools} and \ref{subsec:Software_Publication_Platforms}.

\subsection{Software Description Schemas}
\label{subsec:Software_Description_Schemas}

\paragraph{Software Metadata Standards.}
Metadata schemas (or standards) are sets of metadata elements (or terms) that are compiled to unify the description of artifacts within their scope. Many different metadata schemas exist for a variety of use cases. Whereas \emph{Dublin Core}~\cite{DublinCore} outlines general metadata terms, the \emph{DataCite Schema}~\cite{DataCite} focuses on describing research data. \emph{schema.org} is intended to describe web pages with structured data markups but it is also widely used for other purposes~\cite{zeng_metadata_2022}.

With respect to research software, CodeMeta is a popular community-driven metadata standard. It is based on \emph{schema.org}, which it augments with several additional terms. Various crosswalks exist - that is, mappings from one schema to another - between CodeMeta and other metadata schemas. CodeMeta covers many aspects of software metadata, with some terms focusing on technical details such as file size or operating system and others on administrative information like licenses and links to the software repository. It supports the unambiguous assignment of authors, contributors, licenses, and more via Uniform Resource Identifiers (URIs). The purpose of a software can be specified by means of a textual description, application categories, keywords, and a link to a README file or reference publication. Apart from a coarse classification, the declaration of a software's purpose is therefore still far from being readily machine-actionable; that is, without interpreting (or misinterpreting) natural language. Furthermore, CodeMeta does not include terms for specifying the input and output of a software, nor does it include terms for specifying features or methods implemented by a software. Similarly, the \emph{Citation File Format} schema defines general metadata for the citation of software repositories without describing the software's purpose and interface~\cite{druskat_stephan_2021_5171937}.

In the domain of geoscience, Garijo et al. developed the \emph{Software Desciption Ontology}~\cite{SoftwareDescriptionOntology} by extending their own approach, namely \emph{OntoSoft}~\cite{gil_ontosoft_2015}. OntoSoft elements are structured in six categories: identify, understand, execute, do research, get support, and update. The ontology captures technical metadata like programming language and dependencies and descriptive data like name, website, and contributors. The authors added a description of the input and output data also utilizing the \emph{Scientific Variables Ontology} and aligned \emph{OntoSoft} with CodeMeta. The metadata are published to an open knowledge graph~\cite{OntoSoftPortal}. Garijo et al. support the linking to other instances in the semantic web, like Wikidata, the \emph{Scientific Variables Ontology}, and others. Additionally, they developed programs to support researchers in metadata creation and the search for software models~\cite{garijo_okg-soft_2019, MINT}.

In the domain of bioinformatics, Ison et al. developed the metadata schema \emph{biotoolsXSD} for the software registry \emph{bio.tools}~\cite{ison_tools_2016, ison_biotools_2019, BioTools}. The metadata is expressed as an XML schema and contains 55 elements, of which ten are mandatory. The use of the EDAM ontology as value vocabulary is required for elements such as function, input, and output. The metadata schema also contains software-specific elements like programming language, license, and operating system. The use of an ontology is not required for these.

\paragraph{Interface Description Standards.}
In order to increase its technical interoperability and reusability, software can be documented by means of interface description languages. The syntax of such a language enables the formal and programming language-agnostic description of interface functions and their parameters. Well-known representatives include the Web Service Description Language (WSDL)~\cite{WSDL} and Web Application Description Language (WADL)~\cite{WADL}. Both are XML-based specification standards that describe the syntactical elements of web services and, primarily, how to access them. They are utilized to simplify the information exchange in Web 2.0 application development. Whereas WSDL is used in conjunction with SOAP, WADL enables the description of HTTP (and in particular REST-conform) web services. Both languages provide machine-processable descriptions but do not support taxonomy or ontology information for semantic classification. The WSDL and WADL standards were last updated in 2007 and 2009, respectively.

The OpenAPI specification is an interface description language that focuses on REST APIs~\cite{OpenAPI}. By utilizing YAML and JSON, it is both machine-actionable and human-readable. By default, it is used to define the general properties of APIs, such as the version, contact, and license information or server names and addresses. However, it also defines technical aspects, mainly with respect to REST interface functions like the paths to endpoints, HTTP verbs, parameters or response code descriptions. The OpenAPI standard also allows for the annotation of custom properties using a concept called \emph{extensions} or \emph{x-attributes}. These extensions provide a powerful way of describing additional functionality not covered by the standard specification. As an open and non-proprietary state-of-the-art industry standard, the OpenAPI specification is actively maintained and regularly updated.
  
The Web Ontology Language for Web Services (OWL-S) defines ontologies built on top of the Web Ontology Language (OWL) for describing semantic web services on a technical level, making it more powerful but also more complex than regular description languages (i.a., WSDL and WADL)~\cite{OWL-S}. It describes the purpose of services, how they are accessed, and how they function. Although more powerful than comparable description languages, OWL-S is not an `end-all-be-all' solution to service descriptions and requires domain-specific ontologies for describing domain-specific functionality. Furthermore, its focus on semantic web services greatly reduces its legibility; it was last updated in 2004 and is not suited to easy human reading.

The Functional Mock-up Interface~(FMI) is an open-source standard for simulation software interfaces~\cite{FMI}. All simulation models whose interfaces have been designed along the standard become so called functional mock-up units~(FMUs). The standard ensures that all FMUs are compatible with one another and can be executed in combination on the basis of XML and binary files and C~code containing functions, variables, and mathematical formulas. FMI comes with its own documentation standard, namely the FMI Description Schema, which only applies to FMI conform software. It encompasses general information regarding the FMUs such as name, version, author, and license, as well as technical information like model structures, unit, and type definitions. The schema allows structured extensions to the base standard in order to flexibly meet additional requirements. FMI is still actively maintained today and is used in many industrial companies.

\paragraph{Non-standardized Software Description.}
Software is also described on web pages, where the use of specific terms is typically enforced, but without adopting a metadata schema, thereby only establishing uniformity on the web page itself. Schwarz and Lehnhoff~\cite{schwarz_ontological_2019}, for example, describe a catalog of energy co-simulation components. They use a semantic media wiki to collect information on simulators and add descriptions to the simulation interfaces. The elements of the catalog, which can be used for a metadata schema, are not described in greater detail. The open energy modeling initiative~(openmod) includes a list of energy models in their wiki~\cite{openmodWikiOpenModels}. For each of these, administrative and descriptive metadata are listed, such as license, link to a code repository, model class, and other. The descriptive elements include detailed information on the models. The elements are not formalized as metadata schema and controlled vocabularies are used for neither the elements nor the values. The Open Energy Platform~(OEP) introduces framework and model factsheets to describe frameworks and models~\cite{OEPModelFactsheets}. These descriptions have been further developed based on the non-formalized openmod metadata elements.

In addition to the description by means of metadata, software is described in documentation and specification websites, providing guidance for both users and developers (e.g., see ~\cite{PyPSADocumentation}). The design ideas and specific technical elements of software are typically defined along with their underlying algorithms and procedures. Specifications for the API, user manuals, and examples of applications make it possible to correctly utilize the software. Software documentation is predominantly written in natural language and, therefore, is neither machine-actionable nor easily searchable or comparable. Although such documentation provides rich information, it is not typically considered as part of software metadata. \\

It should be noted that existing software metadata schemas do not include technical documentation about interfaces. And although interface description languages are designed to collect this information, they focus on web services and protocols. As a result, the interface information of software that is not provided as a service is primarily published as non-standardized and non-machine-actionable information on web pages, often without any connection to controlled vocabularies or ontologies. For research software, most of which is not provided as a service, there is not yet a suitable schema that enables semantic interface descriptions. However, the near-code structures of interface description languages and the ability to connect some via extensions to established software metadata schemas offer promising foundations.

\subsection{Software Description Tools}
\label{subsec:Software_Description_Tools}

Documentation is generally regarded as an essential component of software development, and yet it is frequently neglected. This is often due to the fact that considerable effort is involved in writing detailed, well-structured, and version-controlled documentation. A recommended means of alleviating this issue is the use of automated documentation tools~\cite{leeTenSimpleRules2018}, which are specifically designed to aid in the process of creating comprehensible and complete documentation for a software project. There are many such tools available, and although the general objective is the same, they differ in their approach, programming language, or input and output formats.

Many of these tools, e.g., Javadoc~\cite{Javadoc} or Perldoc~\cite{Perldoc}, focus on single programming languages and use source code as their main inputs. By parsing the code, they obtain information on defined types and functions and their relationships. Some documentation tools, such as Doxygen~\cite{Doxygen} or the Sphinx plugin Napoleon~\cite{SphinxNapoleon}, are able to extract this kind of information from bare code; other tools, however, rely on code comments in a determinate format. In either case, additional metadata is typically conveyed via comments. This can comprise, for example, a general description or explanation of a function's parameters. Such information is mostly given as free text and is placed into the final documentation without change. MkDocs~\cite{MkDocs} and, in some cases, Sphinx~\cite{Sphinx} constitute an exception by only manually parsing created files, e.g., containing \emph{reStructuredText}. They can, however, both be extended with plugins that automatically generate said text files from code. The output of documentation tools is nicely-formatted documentation pages, typically using HTML or LaTeX. These pages are easily readable and comprehensible to humans, but hardly machine-actionable. Roxygen2~\cite{Roxygen2} also generates intermediate files that are, in theory, machine-actionable, but, due to their custom data format, are limited in their reusability.

In this regard, Swagger~\cite{Swagger} can be distinguished from other tools. Swagger is used primarily for documenting REST APIs and provides a set of distinct but related tools for that. At its core, Swagger utilizes a YAML file standardized in the OpenAPI Specification. This file is machine-actionable and stores all metadata of an API in a structured, hierarchical way. It can be created manually or generated from code, and, when passed to the appropriate Swagger tools, is used to generate a human-readable documentation web page. Unlike many other tools, Swagger does not require specially-formatted comments within the code in order to extract the information. Furthermore, Sphinx can be extended by a plugin to enable support for OpenAPI specification files, which, as implied, makes it possible for Sphinx to generate interface descriptions from OpenAPI compliant YAML. \\

It should be highlighted that software and, therefore, interface documentation can be parsed automatically from source code and many documentation tools are available. However, most of these rely on code comments that are formulated in natural language and which, therefore, are not directly machine-actionable. In this regard, Swagger is an exception, as it centers around a universal, machine-actionable and standardized metadata file, which is suitable for documentation pages as well as automated reuse. Even though Swagger is intended only for documenting REST interfaces, there is no lock-in to individual programming languages. Because of this inherent flexibility, it offers some potential for the development of generic software documentation workflows.

\subsection{Software Publication Platforms}
\label{subsec:Software_Publication_Platforms}

Software can be made discoverable and available for reuse by being published on a variety of software-specialized publication platforms. Thereby, the distinct purposes and objectives of these platforms vary. Although some store the source code of a software in versioned repositories (e.g.,~\cite{GitHub}), in particular to enable its further development, others aim at the distribution and easy integration of mature programs (e.g.,~\cite{Anaconda}). Some platforms serve as registries, indexing large collections of software and making them searchable using detailed metadata (e.g.,~\cite{PyPi}). Others are dedicated to the provision of technical documentation and user guides (e.g.,~\cite{ReadTheDocs}).

Furthermore, most of the software publication platforms differ in the data formats they accept and in the uploading processes they provide. Even when using similar file formats, the required information or information structures vary. Some platforms, such as Github~\cite{GitHub}, Gitlab~\cite{GitLab}, Bitbucket~\cite{Bitbucket}, or Sourceforge~\cite{Sourceforge}, ingest the source code directly without a specific required structure. Others support the inclusion of metadata configuration files. For example, Anaconda Distribution~\cite{Anaconda} requires a YAML file that describes the project. Maven Central~\cite{Maven} requires an XML POM file for storing metadata. Whereas PyPi~\cite{PyPi} requires a TOML file with information about packages, NPM~\cite{NPM} generates a JSON file based on text prompts. Swaggerhub~\cite{SwaggerHub} requires an OpenAPI-conforming interface description file in YAML format, containing function and argument specifications. Like ReadTheDocs~\cite{ReadTheDocs}, some platforms require a software project to have a documentation folder according to a standard. In this specific case, Sphinx or MkDocs can be used in order to generate such a folder. Platforms like Gitbook~\cite{GitBook}, CRAN~\cite{CRAN}, or Github~Pages~\cite{GithubPages} require programming language-specific files for the installation. For example, submitting a project to CRAN requires first creating a TAR.GZ file. Github~Pages~\cite{GithubPages} can store project documentation via HTML files. The OEP~\cite{OEPModelFactsheets}, Open Research Knowledge Graph (ORKG)~\cite{ORKG, Auer.2020,  Stocker2023_01}, or \emph{bio.tools}~\cite{ison_biotools_2019}, for example, require manually filling forms with project data in order to register it. \\

There is no question that publishing platforms are critical to the dissemination, findability, and reusability of research software within and across academic communities. It is advantageous to employ various platforms in parallel to utilize their distinct strengths to increase the impact and transparency of a software. However, as no uniform format for the exchange and subsequent use of software metadata has yet been identified, metadata must often be collected redundantly and adapted to heterogeneous formats and processes, creating the need for a machine-actionable and programming language-agnostic exchange standard.

\section{The DataDesc Ecosystem}
\label{sec:The_DataDesc_Ecosystem}

This section introduces the DataDesc ecosystem. As a central component, the DataDesc schema, which enables the thorough description of software interfaces, is explained in \Cref{subsec:DataDesc_schema}. Then, in \Cref{subsec:DataDesc_Exchange_Format_and_Utilities}, an exchange format as well as assistance tools are presented, enabling the gathering, storage, and reuse of machine-actionable metadata. Finally, procedures that can be used to share metadata on publishing platforms are defined in \Cref{subsec:DataDesc_Publication_Pipelines}. DataDesc has been released with all of its components presented here under the open MIT license on GitHub~\cite{DataDesc}.

\subsection{DataDesc Schema}
\label{subsec:DataDesc_schema}

Metadata schemas often focus on general information provision, which primarily includes the naming of organizations and persons involved in the development process and the technical and licensing conditions under which the software can be obtained and used. By specifying categories and keywords, they also make a valuable contribution to supporting the findability of software. Within these schemas, however, the description of interfaces can only be superficially embedded in general metadata elements. Although this information already provides important insights into a software, it is not sufficient to facilitate its interoperability and reusability in a machine-actionable way. Therefore, the DataDesc schema\footnote{More precisely, DataDesc's schema is a metadata application profile, as it combines term definitions from existing metadata schemas for a particular purpose. However, as this technical difference is not decisive, the more common term \emph{schema} is used.} compensates for this omission by providing a framework for the detailed annotation of individual software interface components, leading to insights into how and with which data and programs a software can be used. The DataDesc schema promotes the reuse and integration of research software and, thereby, is an ideal extension to existing metadata schemas.

\begin{figure}
	\centering
	\includegraphics[width=1.0\linewidth]{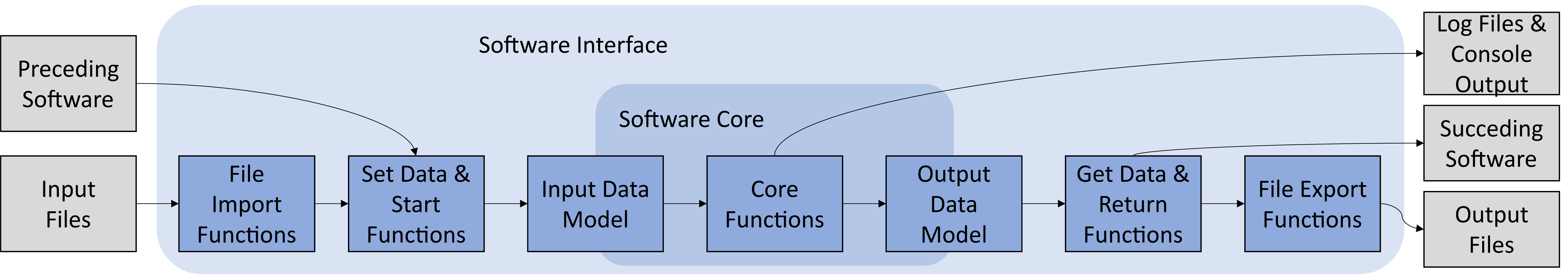}
	\caption{Schematic representation of the generic information flow between software interface and core components.}
	\label{fig:InterfaceSchematic}
\end{figure}

An interface, as schematically depicted in \Cref{fig:InterfaceSchematic}, serves as a connection point for users and programs to interact with a software. It is composed of the functions through which data can be inputted into and retrieved from the software. These functions are distinguished from the inner functions, which form the logic of the software core. The program core can only be addressed indirectly via the interface, whereby the structures and formats of the information flow are defined by the interface functions and internal data models.

\begin{figure}	
	\centering
	\includegraphics[width=1.0\linewidth]{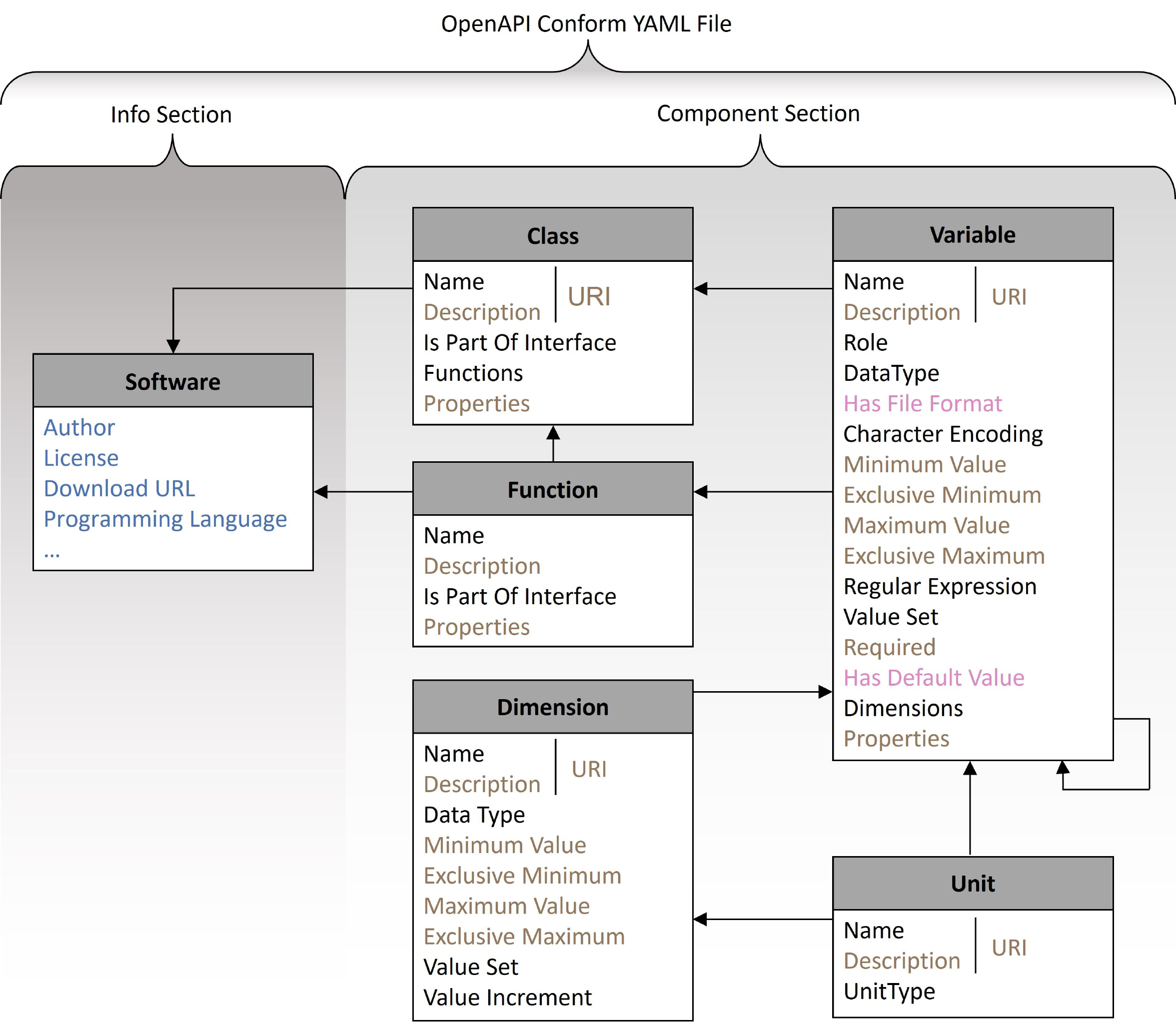}
	\caption{Structure and content of the DataDesc schema within an OpenAPI-conforming YAML file. While a software is described with \emph{CodeMeta} (blue) in the info section, its objects are described in the component section with \emph{DataDesc} (black), which re-uses terms from \emph{OpenAPI} (brown) and the \emph{Software Description Ontology} (pink). All objects are arranged in a hierarchical structure, which is indicated by arrows pointing from child to parent objects.}
	\label{fig:DataDescSchema}
\end{figure}

An interface description performed with the DataDesc schema formally identifies the characteristics of an interface according to a collection of metadata elements, detailed in \Cref{fig:DataDescSchema}, whose meaning and use are explained in the following. The schema comprises the naming~(\emph{Name}) and description~(\emph{Description}) of all functions, which are part of an interface~(\emph{Is Part Of Interface}) and over which a software can be addressed. In order to enable easy and in particular error-free use of a software, the functions' parameters, as well as their underlying data models, must be described in detail. To adequately characterize variables serving as the input or output parameters of interface functions~(\emph{Role}), their intended and allowed data must be described in terms of contents, formats, values, and structures.

\paragraph{Data Content Description.}
In order to digitally process information, it must be stored in the form of variables. In the course of software development, the data content of each variable is defined. This refers explicitly to the referencing of real world concepts, such as the height or weight of a person, and not of data types, which specify whether variables can contain \emph{integers}, \emph{floats}, \emph{strings}, or similar. A precise understanding of the meaning of the data content a software requires, processes and outputs is essential for its correct and intended use.

However, capturing meanings is not a trivial task. Depending on the demand for precision and generality, describing data content involves varying degrees of effort. The easiest approach is to sensibly name variables during software development~(\emph{Name}) and explain them further in docstrings~(\emph{Description}). However, these names and free text descriptions almost always leave considerable room for interpretation as to the meaning of the data content. Instead, it is more interoperable to reference concepts from ontologies (with their respective \emph{URIs}), which often provide unambiguous definitions that are agreed upon in the respective research domain~\cite{heiler1995semantic}. Of course, the open collaborative development of such concepts with the broadest possible participation and agreement within a domain is a labor-intensive process requiring well-organized community infrastructures~\cite{booshehri2021introducing}.

If the variable is numerical, documenting the meaning alone is insufficient for fully describing its data content. In this case, additional information about a unit is necessary, so that, for example, a duration of seven hours can be distinguished from one of seven seconds. Just as with the concepts before, a unit can be specified by a name~(\emph{Name}) and a description~(\emph{Description}), or an ontology reference~(\emph{URI}).

In the context of software interfaces, specific concepts and units need not always be declared. In order to enable a greater degree of freedom in data entry and so to enable a more flexible application of a software, intended data contents can be more broadly indicated. For example, specifying the general concept \emph{means of transport} indicates that the software can process data about \emph{bicycles}, \emph{trains}, \emph{cars}, and so forth. Likewise, instead of specific units such as \emph{meters}, \emph{centimeters} or \emph{miles}, the unit type \emph{length} can be used~(\emph{Unit Type}). In this context, the use of ontologies offers the advantage that they often already include information that specifically relates to more general concepts.

\paragraph{Data Format Description.}
The format of a variable defines how the information it contains is to be encoded into binary data and subsequently interpreted. It provides information about which operations may be applied to the data content. The format is defined by the data type of a variable~(\emph{Data Type}). There are primitive and complex data types that can be native to programming languages or which come as custom data types provided by libraries. Primitive data types, such as \emph{strings}, \emph{integers}, or \emph{booleans}, can hold single values. Complex data types, like \emph{lists}, \emph{tables}, \emph{arrays}, or \emph{classes}, can group multiple instances of primitive data types.

\begin{figure}
	\centering
	\includegraphics[width=1.0\linewidth]{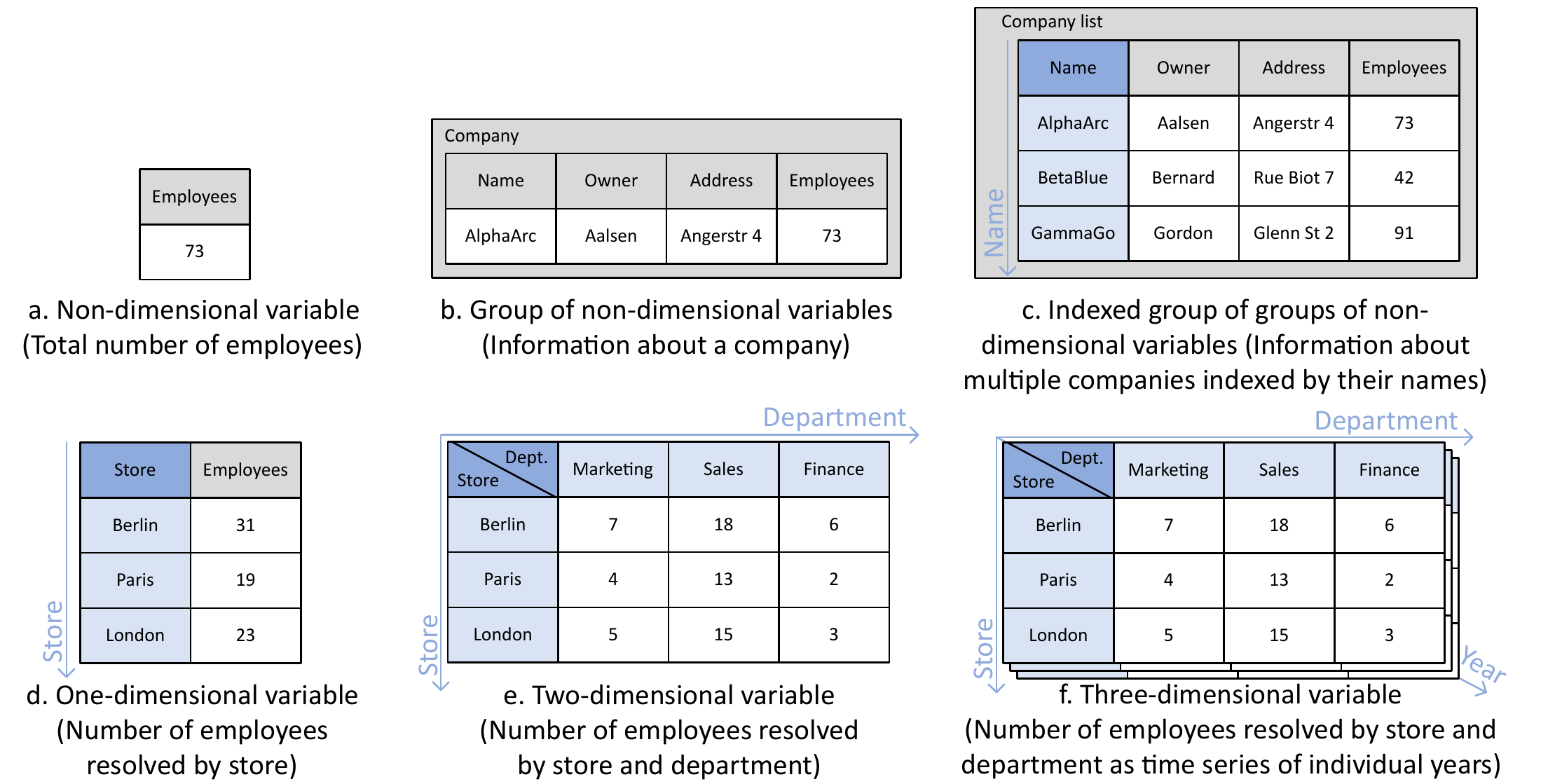}
	\caption{Comparison of widely-used data structures based on an example of information about companies. Variables are shown in gray and their values in white, whereas dimensions are displayed in dark blue with their indices in light blue.}
	\label{fig:DataStructureComparison}
\end{figure}

As complex data types can also recursively contain complex data types, nested structures of arbitrary depth and complexity are possible, although their final level can only contain primitives. Complex structures of this kind are often used to define data models, which summarize the input and output data of research software into single data objects and make them centrally accessible (e.g.,~\cite{FINEsEnergySystemModel},~\cite{IAMCsPyamDataModel}). Oftentimes, classes are used at the highest level for the representation of such data models, to which the interface functions~(\emph{Functions}) for importing and storing, as well as for reading out and exporting, are also assigned (cf. \Cref{fig:InterfaceSchematic}). In order to describe not only the data types of the function parameters but also the formats nested within them, hierarchical data formats are mapped in the DataDesc schema by means of hierarchical parent-child relationships. To avoid redundant descriptions of complex data models with each function parameter based on them, the DataDesc schema also offers the option of creating separate reusable descriptions of data model classes that can also be referenced~(\emph{Name}, \emph{Description}, \emph{URI}, \emph{Is Part Of Interface}). \Cref{fig:DataStructureComparison}~(a.-c.) exemplifies how four variables of the primitive data types of \emph{integer} and \emph{string} are grouped (e.g., in the format of a \emph{class}). Different instantiations of this class are further grouped (e.g., in a \emph{dictionary}).

Files indirectly represent another complex data type, as they can also contain and group data of arbitrary types. As is shown in \Cref{fig:InterfaceSchematic}, reading in files is a widely used method of transferring data to a research software, which is why an interface description must also inform regarding permitted file formats that can be processed without errors. For each variable containing a file reference, whatever the variable itself, e.g., file type \emph{string} or \emph{file object}, DataDesc gives the option of specifying the format, e.g., text formats like \emph{XML}, \emph{HTML}, or \emph{TEXT} or binary formats like \emph{PDF}, \emph{XLS}, or \emph{JPG} of a referenced file~(\emph{Has File Format}). Beyond that, the character encoding of text formats, e.g., \emph{UTF-8}, \emph{ASCII}, or \emph{ISO 8859-1}, can be specified to support the correct interpretation of text data~(\emph{Character Encoding}).

\paragraph{Data Value Description.}
When creating software, a permissible value range must be defined for each variable, guaranteeing the technically error-free processing and consistency of content for all values from within this range. Technically, the value range is defined in many programming languages by the choice of a variable type. In Java, for example, a variable of type \emph{float} allows all floating point values from $-3.4 * 10^{38}$ to $3.4 * 10^{38}$, whereas a \emph{boolean} can only accept the values of \emph{true} and \emph{false}.

In addition, a value range can be further limited based on content considerations. For example, if a longitude is to be stored in a \emph{float} variable, only floating point values from $-180$ to $+180$ degrees should be considered valid~(\emph{Minimum Value}, \emph{Exclusive Minimum}, \emph{Maximum Value}, and \emph{Exclusive Maximum}). If text variables are only allowed to contain certain patterns, this can be defined through \emph{regular expressions} (\emph{Regular Expression}). For example, if a filename is to consist of only letters, numbers, and underscores, this can be defined using the expression, \verb/^[A-Za-z0-9_]+$/. If the allowed value range of a variable should be fully restricted to predefined values, e.g., \emph{North}, \emph{East}, \emph{South}, and \emph{West}, DataDesc schema offers the possibility of documenting them in the form of value sets~(\emph{Value Set}).

It is also part of the value description to specify whether a variable is an optional parameter~(\emph{Required}). If this is the case, it does not need to be set when the respective function is called upon. In this context, a default value can also be specified, and is used if the variable is not explicitly set (\emph{Has Default Value}).

\paragraph{Data Structure Description.}
For variables of complex data types, the internal data structures must be described at both a technical and contextual level so as to enable the correct accessing of individual values and the determination of their respective meanings. The processing procedures of software programs are designed on the basis of specific data structures whose declaration is the task of interface descriptions. The correct function of a software is not ensured if the structure of the passed data differs from the intended data structure.

\Cref{fig:DataStructureComparison} (b.) shows four independent variables which, as they represent information characterizing the same single object, are combined in a grouping variable \emph{company}, which itself must be described. The technical structuring of the data thereby maps their context and relates the four variables to each other and to the grouping variable: \emph{Number of employees} becomes \emph{Number of employees of the company AlphaArc}. In order to capture this kind of context within an interface description, the DataDesc schema utilizes hierarchical parent-child structures to map relationships between variables. For this purpose, variables that are the child objects of data model classes or grouping variables are listed as their properties~(\emph{Properties}).

In addition to grouping, the dimensional resolution of information represents a significant structural pattern. \Cref{fig:DataStructureComparison}~(a., d., and e.) shows the increasing resolution of the initially non-dimensional information: \emph{the AlphaArc company has a total of 73 employees}. This information does not change subsequently, but the single value is broken down into individual values per store and then per store and department. Each dimension along which the information is resolved is listed as part of the variable~(\emph{Dimensions}). The DataDesc schema allows individual specification of the meaning~(\emph{Name}, \emph{Description}, \emph{URI}), index type~(\emph{Data Type}), and index range~(\emph{Has Minimum Value}, \emph{Has Maximum Value}, \emph{Value Set}, and \emph{Value Increment}) of the dimensions. In this context, the combination of dimension indexes is unique, which is why it acts as a key and enables the unique identification and retrieval of each individual value. At the same time, an individual context is defined for each value: \emph{15 employees is the team size of the sales department in the London store}, for example.

The structural description provides not only information about the context of values but also about data access mechanisms that might be expected by interface functions (see \Cref{fig:DataStructureComparison} (c.)). For the structure of the \emph{company list}, which can be, e.g., in the form of a \emph{Python dictionary}, \emph{pandas DataFrame}, \emph{Java HashMap}, or \emph{SQL table}, the variable \emph{Name} was determined as a key index due to the identifying character of its values. As a dimension of the \emph{company list}, the variable \emph{Name} allows access to individual datasets.

\Cref{fig:DataStructureComparison} (f.) shows another structure that combines grouping and resolution by adding the third dimension \emph{year} to the resolution of the \emph{number of employees} based on the two dimensions of \emph{store} and \emph{department}. Here, the total number of 73 employees is not broken down further, but put in the context of a specific year, e.g., \emph{15 employees is the team size of the sales department in the London store in the year 2010}. Together with uniform information for, e.g., the years 2015 and 2020, this third dimension turns the dataset into a time series.

\subsection{DataDesc Exchange Format and Utilities}
\label{subsec:DataDesc_Exchange_Format_and_Utilities}

In addition to the schema for describing interfaces, the exchange format for the integrated storage and flexible subsequent use of software metadata represents the second core component of the DataDesc ecosystem. The OpenAPI specification was chosen as its foundation, as it allows a programming language-agnostic description of software that is usable by both humans and computers. A software is described in a single OpenAPI-conforming YAML file. Its basic structure consists of a hierarchical object tree, that is subdivided into the two sections of \emph{info} and \emph{components} (compare \Cref{fig:DataDescSchema}). In the \emph{info} section, all general information, for example along a general software metadata schema such as CodeMeta, can be accommodated. If metadata elements are required for this that are not provided for in the OpenAPI specification, they can be added by means of \emph{x-attribute} extensions without violating the standard. In the \emph{components} section, the technical interface metadata, as described by the DataDesc schema, for example, can be specified. The \emph{x-attributes} again provide the opportunity to compensate for missing OpenAPI metadata elements. In addition, they form the basis for using the standard not only to describe REST-compliant interfaces; they can also be used to arrange and annotate code elements such as classes, functions, and parameters in the hierarchical object tree according to individual software interface designs.

In order to support the description of software based on its general properties and the transfer of this information into the exchange format chosen in the DataDesc approach, a \emph{browser-based input form} was added to the ecosystem. The metadata fields of the form thereby map to the CodeMeta standard, as this is already widely used and can be applied across research domains. Thus, the uncomplicated input tool is an alternative to the JSON-based CodeMeta generator~\cite{CodeMetaGenerator} and fits seamlessly into the DataDesc approach.

Unlike the general metadata, the technical documentation is produced directly in the source code of a program, which is why the definition of a machine-actionable formatting of this information, as well as its automated parsing and transfer into the exchange format must be made individually for each programming language. In the context of Python software, code components related to the interface are supplemented by decorators and individually marked up by means of DataDesc schema elements. A \emph{Python parser} specifically developed for this schema extracts both the relevant code structures and their metadata and automatically generates the hierarchical object tree from them, which is then stored in an exchange format-compliant file.

The DataDesc utilities are complemented by a \emph{tool for merging} the DataDesc files, so that the entire description of a software can be represented in a single concise documentation file that is easily exchangeable. Its OpenAPI conformity also ensures high interoperability, as a multitude of publicly-available tools can be applied to it~\cite{Swagger, OpenAPITools}.

\subsection{DataDesc Publication Pipelines} 
\label{subsec:DataDesc_Publication_Pipelines}

Making it possible for developers to create software metadata and documentation only once and then flexibly reuse it is one of the main objectives of the DataDesc approach. Against this background, technical processes are defined and, where necessary, supported with scripts that enable the collected information to be disseminated on software publication platforms. These publication pipelines are unique to each platform and subject to automation. To upload data to any of the publication platforms mentioned below, a free user account must first be created.

The OpenAPI-conforming YAML file can be uploaded and published on SwaggerHub~\cite{SwaggerHub} via its GUI or API, without any need for modifications. To publish the description on GitHub~\cite{GitHub}, it is sufficient to add the file to the software's versioned repository and reference it in the central README file. To make the documentation more visually-appealing, a link to a SwaggerHub-hosted documentation page can be included. The registration of Python-based software and its metadata in the Python Package Index (PyPi)~\cite{PyPi} has been fully automated. With a DataDesc script utilizing restructuring and conversion tools, the YAML file and corresponding software source code are reformatted, uploaded, and published on the platform. The publication on ReadTheDocs~\cite{ReadTheDocs} can also ingest information based on a DataDesc exchange file. In order to upload the documentation in the appropriate format, it must first be created using, for example, Sphinx with its extension for the parsing of OpenAPI specifications. Then, a GitHub repository comprising the generated documentation can be imported. The ORKG~\cite{ORKG, Auer.2020, Stocker2023_01} provides a GUI and an API for uploading software metadata, which can be entered manually into a form. In addition, a script was added to the DataDesc ecosystem to automate the translation of the exchange format into the ORKG template structures~\cite{Karras.2021}. Currently, this mapping must still be performed individually by each user. However, work is underway to include this functionality in the ORKG. As part of the development of the Open Energy Research Graph~\cite{OEKG}, efforts are being made to ensure that the exchange format can also be processed directly within the OEP~\cite{OEPModelFactsheets}.

\section{Application Case}
\label{sec:Application_Case}

In this section, the application of the DataDesc approach to a research software is discussed. For this, the open source FINE framework~\cite{WELDER20181130} from the \emph{Energy Transformation Pathway Optimization Suite} (ETHOS) was chosen to illustrate a use case that is both realistic in terms of complexity and intriguing with respect to the interfaces provided. Using selected excerpts from the created interface documentation shown in \Cref{fig:YAMLExcerpt}, the OpenAPI-compliant syntax of the YAML file generated using the DataDesc utilities is presented and the semantic expression capabilities of the DataDesc schema are assessed. To allow for a more in-depth review of the entire DataDesc approach, all code and documentation files created as part of this example application are published in the DataDesc repository~\cite{DataDesc}. \\

FINE is a Python package with a five-year development history originating in the research domain of energy systems analysis~\cite{WELDER20181130}. It enables the modeling, optimization, and evaluation of energy systems. In addition to accounting for technical and environmental constraints, its optimization also seeks to minimize total annual energy system costs. It supports the creation and computation of spatially, temporally, and technologically highly-resolved models while integrating complexity-reduction techniques to shorten computation times. In its current version, 2.2.2 from 2022, the framework includes around 20,000 lines of code (excluding blank lines) and 10,000 lines of code documentation. Although the source code of the software project is hosted on GitHub~\cite{FINERepository}, the user documentation is published on ReadTheDocs~\cite{FINEDocumentation}. The documentation pages are based on the inline docstrings in the source code and were automatically generated using the Sphinx package. In addition, a short entry in the OEP's software framework list was written for the framework~\cite{FINEOEPModelFactsheet}.

The FINE software is based on a central data model, namely the \emph{Energy System Model}~(ESM), which is represented by the ESM and component classes. It holds multidimensional information pertaining to the energy system under investigation and comprises all characteristics of its components, e.g., for the provision, transmission, and storage of energy. As input, besides basic parameters for calculation control, the software requires the general conditions of the energy system and the techno-economic parameters of its components. As output, it provides information for the design and operation of a minimum-cost energy system. As the ESM incorporates all of this data, it simultaneously serves as the software's input and output data model.

As noted in the general schematic in \Cref{fig:InterfaceSchematic}, the FINE interface offers the possibility of reading the input data from files or having them transferred by preceding software. However, in the second case the information can be gathered step by step in the data model classes; for file-based information transfer, a single complex file containing all parameters must be provided. Both Excel and NetCDF file formats are accepted for this purpose. On the output side of the interface, the result data can also be saved in Excel or NetCDF form or visually depicted using a range of plotting functions. \\

\begin{figure}
	\centering
	\footnotesize
	\definecolor{dark_gray}{rgb}{0.3,0.3,0.3}
	\usemintedstyle{default}
	\begin{minted}[
        gobble=4,
        frame=single,
        escapeinside=||,
        linenos
        ]{YAML}
openapi: 3.0.0
info: ### Info section
  title: FINE - A Framework for Integrated Energy System Assessment
  version: 2.2.2
  x-first-release: '2018-11-12'
  x-programming-lang: Python
components: ### Components section
  Component: ### FINE Component class
    description: The Component class includes...
    properties: ### Class Variables
      capacityMax:
        x-dimensions: &id001 ### Referencable inner Pandas Dataframe structure
          location:
            ItemMinimumValue: 0
            UnitType: spatial identifier
          time:
            ItemMinimumValue: 0
            UnitType: temporal identifier
  EnergySystemModel: ### FINE Energy System Model class
    properties: ### Class Variables
      numberOfTimeSteps:
        type: integer
        x-DefaultValue: 8760
        x-MinimumValue: 0 ### Allowed value range
        x-ExclusiveMinimum: true
    required: ### List of required parameters
    - numberOfTimeSteps
    x-functions: ### Class functions
      aggregateTemporally:
        properties: ### Function parameters
          clusterMethod:
            x-ValueSet: ### Allowed value set
            - averaging
            - k_means
            x-VariableRole: input ### Input parameter
      removeComponent:
        return: ### Return value
          $ref: '#/components/schemas/Component' ### Referencing data structure
      readNetCDFtoEnergySystemModel:
        properties: ### Function parameters
          filePath: ### Path to external file
            type: string
            x-FileFormat: NetCDF
            x-NetCDFFolders: ### Inner structure of referenced NetCDF file
              Input Data: ### Data folders
              - Conversion
              - Storage
              Parameters: ### Control parameters
              - numberOfTimeSteps
              - verboseLogLevel
    \end{minted}
	\caption{Compilation of selected lines of code from the YAML file generated to document the FINE interfaces to represent the OpenAPI-compliant syntax and its semantics within the DataDesc schema.}
	\label{fig:YAMLExcerpt}
\end{figure}

The FINE interfaces for reading Excel and NetCDF files are implemented by one function each, which mainly obtain a path to the respective input file. Here, DataDesc offers the possibility, in addition to the superficial description of the string variables, of going into depth and also describing the necessary internal structures of the input files (cf. \Cref{fig:YAMLExcerpt}, lines 39-50). Thus, for the NetCDF interface, the control parameters and input variables were documented in the clearly structured hierarchical data format, which arranges the information according to entity types and in each case lists their attributes in accordance with their different dimensional characteristics. The documentation of the Excel data structure required more effort, as it does not group the data by entities, but distributes them to different spreadsheets depending on their dimensional resolutions. The resulting tables, in which the multi-dimensional attribute characteristics of different entities are mapped by means of different index columns, required precise documentation to define the boundaries between individual datasets. In both cases, documentation could be created to help users understand the given file structures and arrange their own input data accordingly. The documentation effort depended on the straightness and non-ambiguity of the data model structures.

For the documentation of the programmatic interface, the constructors of the ESM and the component classes were described using DataDesc. Here, the use of Python's dataclass decorators and minor code adjustments to the constructors enabled the transformation of previously free-text docstrings into unambiguous key-value pairs. Variable names, comments, types, roles, and default values could easily be mapped to the DataDesc annotation syntax (line 7 ff). For Python native types, the variable types were annotated directly into the code using type hints. For custom types, such as Panda's dataframes, these annotations, including more detailed structural information, were written in the decorators. In addition, the value range constraints that some variables are subject to could also be integrated into the documentation with the metadata elements contained in the DataDesc schema (lines 24-25). Furthermore, the permitted value sets and also complex data structures of the FINE variables could be described in detail (lines 12-18, 32-34). To minimize the documentation effort, value sets and data structures that apply to several variables were documented only once and then referenced repeatedly (lines 12 and 38).

Content dependencies, like the \emph{costUnit} parameter of the ESM class, which determines the currency and units of all monetary variables, can also be expressed through referencing. The description of procedural dependencies, in which the value of a variable influences the software-internal calculation processes, must be represented so far as free text comments. An example for this is the component class that contains the Boolean parameter \emph{hasCapacityVariable}, which, if set to \emph{true}, causes the \emph{capacityVariableDomain} and \emph{capacityPerPlantUnit} variables to be ignored in the calculations. Work is currently underway to formally integrate this form of dependencies into the schema. Another dependency type results from the fact that FINE integrates the \emph{tsam}~\cite{tsamRepository, HOFFMANN2021117825} library for the purpose of temporal data aggregation and partially maps the external interface within its own interface. In a function of the ESM class, for example, the parameter aggregation method can be selected (lines 31-34). The permitted value set, which includes options like \emph{averaging} or \emph{k-means}, is specified by the external library and manually included in the FINE documentation. In the future, the documentation of independent programs could be integrated and reused automatically if they are also made machine-actionable by means of the DataDesc schema.

\section{Summary and Conclusions}
\label{sec:Summary_and_Conclusions}

The FAIR principles and their adaptations to research software have received much attention and support. To effectively reuse a software, the software itself and its interfaces must be clearly defined and made understandable, ideally in a machine-actionable manner. However, most research software today is not documented or published in a way that provides detailed and machine-actionable interface descriptions. Instead, software metadata is often focused on the compact provision of general information, whereas documentation pages, including detailed, technical information are primarily in natural language and not machine-actionable.

Therefore, the DataDesc ecosystem was presented in this article as an approach to describing the data models of software interfaces using detailed and machine-actionable metadata and as an extension to existing research software metadata. In pointing out that there must be a differentiation between data structures and data formats, it was shown how to consistently describe data structures, and by this support the interoperability of software to other programs and data files. In addition to a specialized metadata schema, an exchange format and support tools for the easy collection and automated publishing of software documentation were introduced. Using the FINE framework as an example, the practical applicability of DataDesc and its limitations were shown. It is hoped that DataDesc will facilitate the description of software interfaces with detailed and machine-actionable metadata enough to make it common practice, leading to increased interoperability, findability, and, therefore, reusability of research software.

In future research, DataDesc is to be enhanced and applied to the description of datasets. Furthermore, extending DataDesc to programming languages beyond Python would be an interesting research direction. With both software interfaces and data being described with DataDesc, the foundation for automatically composing and executing computational workflows has been laid, which will hopefully increase the reuse of research software and the reproducibility of computational research
in the future.

\section{Credit statement}

Conceptualization: P.K., J.G., O.K., and S.F.; methodology: P.K., J.G., O.K., S.F., D.N., J.S., R.P., and F.E.; software: D.N., J.S., and O.K.; validation: R.P., P.K.; investigation: P.K., J.G., O.K., D.N., J.S., R.P., and S.F.; data curation: J.S.; writing – original draft: P.K., J.G., O.K., D.N., J.S., R.P. and S.F.; writing - review
\& editing: P.K., J.G., O.K., D.N., J.S., R.P., S.F., F.E., N.P., J.W., A.N., S.A., and D.S.; visualization: P.K., J.G.; supervision: D.S., S.A., and A.N.; project administration: P.K., O.K.; funding acquisition: D.S., S.A., and A.N.

\section{Acknowledgements}
\label{sec:Acknowledgements}

The authors would like to thank the Federal Ministry for Economic Affairs and Energy of Germany (BMWi) for supporting this work with a grant for the project LOD-GEOSS (03EI1005B).

Furthermore, the authors are grateful to the German Federal Government, the German State Governments, and the Joint Science Conference (GWK) for their funding and support as part of the NFDI4Ing and the NFDI4Energy consortia. Funded by the German Research Foundation (DFG) – 442146713; 501865131.

In addition, the work was supported by the Lower Saxony Ministry of Science and Culture within the Lower Saxony “Vorab” of the Volkswagen Foundation under Grant 11-76251-13-3/19–ZN3488 (ZLE), and by the Center for Digital
Innovation (ZDIN).

This work was also supported by the Helmholtz Association as part of the program “Energy System Design”.

\printbibliography

@inproceedings{Karras.2021,
    title={{Researcher or Crowd Member? Why not both! The Open Research Knowledge Graph for Applying and Communicating CrowdRE Research}},
    author={Karras, Oliver and Groen, Eduard C and Khan, Javed Ali and Auer, S{\"o}ren},
    booktitle={IEEE 29th International Requirements Engineering Conference Workshops},
    year={2021},
    organization={IEEE}
}

@article{Stocker2023_01,
    title = {FAIR Scientific Information with the Open Research Knowledge Graph},
    author = {Markus Stocker and Allard Oelen and Mohamad Yaser Jaradeh and Muhammad Haris and Omar Arab Oghli and Golsa Heidari and Hassan Hussein and Anna-Lena Lorenz and Salomon Kabenamuala and Kheir Eddine Farfar and Manuel Prinz and Oliver Karras and Jennifer D'Souza and Sören Auer},
    doi = {10.3233/FC-221513},
    year = {2023},
    date = {2023-01-11},
    urldate = {2023-01-11},
    journal = {FAIR Connect},
    volume = {1},
    number = {1},
    publisher = {IOS Press}
}

@misc{druskat_stephan_2021_5171937,
    author = {Druskat, Stephan and Spaaks, Jurriaan H. and Chue Hong, Neil and Haines, Robert and Baker, James and Bliven, Spencer and Willighagen, Egon and Pérez-Suárez, David and Konovalov, Alexander},
    title = {Citation File Format},
    month = aug,
    year = 2021,
    publisher = {Zenodo},
    version = {1.2.0},
    doi = {10.5281/zenodo.5171937},
    url = {https://doi.org/10.5281/zenodo.5171937}
}

@article{Auer.2020,
    title = {{Improving Access to Scientific Literature with Knowledge Graphs}},
    author = {Sören Auer and Allard Oelen and Muhammad Haris and Markus Stocker and Jennifer D’Souza and Kheir Eddine Farfar and Lars Vogt and Manuel Prinz and Vitalis Wiens and Mohamad Yaser Jaradeh},
    volume = {44},
    number = {3},
    journal = {Bibliothek Forschung und Praxis},
    year = {2020}
}

@article{wilkinsonFAIRGuidingPrinciples2016,
	title = {The {FAIR} {Guiding} {Principles} for scientific data management and stewardship},
	volume = {3},
	issn = {2052-4463},
	number = {1},
	journal = {Scientific data},
	author = {Wilkinson, Mark D and Dumontier, Michel and Aalbersberg, IJsbrand Jan and Appleton, Gabrielle and Axton, Myles and Baak, Arie and Blomberg, Niklas and Boiten, Jan-Willem and da Silva Santos, Luiz Bonino and Bourne, Philip E},
	year = {2016},
	pages = {1--9},
}

@article{lamprecht_towards_2020,
	title = {Towards {FAIR} principles for research software},
	volume = {3},
	issn = {2451-8484},
	url = {https://content.iospress.com/articles/data-science/ds190026},
	doi = {10.3233/DS-190026},
	language = {en},
	number = {1},
	urldate = {2021-01-20},
	journal = {Data Science},
	author = {Lamprecht, Anna-Lena and Garcia, Leyla and Kuzak, Mateusz and Martinez, Carlos and Arcila, Ricardo and Martin Del Pico, Eva and Dominguez Del Angel, Victoria and van de Sandt, Stephanie and Ison, Jon and Martinez, Paula Andrea and McQuilton, Peter and Valencia, Alfonso and Harrow, Jennifer and Psomopoulos, Fotis and Gelpi, Josep Ll and Chue Hong, Neil and Goble, Carole and Capella-Gutierrez, Salvador},
	month = jan,
	year = {2020},
	pages = {37--59},
}

@article{hasselbring_fair_2020,
	title = {From {FAIR} research data toward {FAIR} and open research software},
	volume = {62},
	issn = {1611-2776, 2196-7032},
	url = {https://www.degruyter.com/view/journals/itit/62/1/article-p39.xml},
	doi = {10.1515/itit-2019-0040},
	language = {en},
	number = {1},
	urldate = {2020-12-02},
	journal = {it - Information Technology},
	author = {Hasselbring, Wilhelm and Carr, Leslie and Hettrick, Simon and Packer, Heather and Tiropanis, Thanassis},
	month = feb,
	year = {2020},
	pages = {39--47},
}

@article{katz_taking_2021,
	title = {Taking a fresh look at {FAIR} for research software},
	volume = {2},
	issn = {2666-3899},
	url = {https://www.cell.com/patterns/abstract/S2666-3899(21)00036-2},
	doi = {10.1016/j.patter.2021.100222},
	language = {English},
	number = {3},
	urldate = {2021-04-21},
	journal = {Patterns},
	author = {Katz, Daniel S. and Gruenpeter, Morane and Honeyman, Tom},
	month = mar,
	year = {2021},
}

@inproceedings{schwarz_ontological_2019,
	address = {Vienna, Austria},
	title = {Ontological {Integration} of {Semantics} and {Domain} {Knowledge} in {Energy} {Scenario} {Co}-simulation},
	isbn = {978-989-758-382-7},
	shorttitle = {Ontological {Integration} of {Semantics} and {Domain} {Knowledge} in {Energy} {Scenario} {Co}-simulation},
	doi = {10.5220/0008069801270136},
	language = {en},
	booktitle = {Proceedings of the 11th {International} {Joint} {Conference} on {Knowledge} {Discovery}, {Knowledge} {Engineering} and {Knowledge} {Management}},
	publisher = {SCITEPRESS - Science and Technology Publications},
	author = {Schwarz, Jan and Lehnhoff, Sebastian},
	year = {2019},
	pages = {127--136},
}

@article{ison_biotools_2019,
	title = {The bio.tools registry of software tools and data resources for the life sciences},
	volume = {20},
	issn = {1474-760X},
	url = {https://doi.org/10.1186/s13059-019-1772-6},
	doi = {10.1186/s13059-019-1772-6},
	number = {1},
	urldate = {2021-02-05},
	journal = {Genome Biology},
	author = {Ison, Jon and Ienasescu, Hans and Chmura, Piotr and Rydza, Emil and Ménager, Hervé and Kalaš, Matúš and Schwämmle, Veit and Grüning, Björn and Beard, Niall and Lopez, Rodrigo and Duvaud, Severine and Stockinger, Heinz and Persson, Bengt and Vařeková, Radka Svobodová and Raček, Tomáš and Vondrášek, Jiří and Peterson, Hedi and Salumets, Ahto and Jonassen, Inge and Hooft, Rob and Nyrönen, Tommi and Valencia, Alfonso and Capella, Salvador and Gelpí, Josep and Zambelli, Federico and Savakis, Babis and Leskošek, Brane and Rapacki, Kristoffer and Blanchet, Christophe and Jimenez, Rafael and Oliveira, Arlindo and Vriend, Gert and Collin, Olivier and van Helden, Jacques and Løngreen, Peter and Brunak, Søren},
	month = aug,
	year = {2019},
	pages = {164},
}

@inproceedings{gil_ontosoft_2015,
	address = {New York, NY, USA},
	series = {K-{CAP} 2015},
	title = {{OntoSoft}: {Capturing} {Scientific} {Software} {Metadata}},
	isbn = {978-1-4503-3849-3},
	shorttitle = {OntoSoft},
	doi = {10.1145/2815833.2816955},
	urldate = {2021-06-22},
	booktitle = {Proceedings of the 8th {International} {Conference} on {Knowledge} {Capture}},
	publisher = {Association for Computing Machinery},
	author = {Gil, Yolanda and Ratnakar, Varun and Garijo, Daniel},
	month = oct,
	year = {2015},
	pages = {1--4},
}

@inproceedings{garijo_okg-soft_2019,
	title = {{OKG}-{Soft}: {An} {Open} {Knowledge} {Graph} with {Machine} {Readable} {Scientific} {Software} {Metadata}},
	shorttitle = {{OKG}-{Soft}},
	doi = {10.1109/eScience.2019.00046},
	booktitle = {2019 15th {International} {Conference} on {eScience} ({eScience})},
	author = {Garijo, Daniel and Osorio, Maximiliano and Khider, Deborah and Ratnakar, Varun and Gil, Yolanda},
	month = sep,
	year = {2019},
	pages = {349--358},
}

@article{ison_tools_2016,
	title = {Tools and data services registry: a community effort to document bioinformatics resources},
	volume = {44},
	issn = {0305-1048},
	shorttitle = {Tools and data services registry},
	url = {https://doi.org/10.1093/nar/gkv1116},
	doi = {10.1093/nar/gkv1116},
	number = {D1},
	urldate = {2021-04-14},
	journal = {Nucleic Acids Research},
	author = {Ison, Jon and Rapacki, Kristoffer and Ménager, Hervé and Kalaš, Matúš and Rydza, Emil and Chmura, Piotr and Anthon, Christian and Beard, Niall and Berka, Karel and Bolser, Dan and Booth, Tim and Bretaudeau, Anthony and Brezovsky, Jan and Casadio, Rita and Cesareni, Gianni and Coppens, Frederik and Cornell, Michael and Cuccuru, Gianmauro and Davidsen, Kristian and Vedova, Gianluca Della and Dogan, Tunca and Doppelt-Azeroual, Olivia and Emery, Laura and Gasteiger, Elisabeth and Gatter, Thomas and Goldberg, Tatyana and Grosjean, Marie and Grüning, Björn and Helmer-Citterich, Manuela and Ienasescu, Hans and Ioannidis, Vassilios and Jespersen, Martin Closter and Jimenez, Rafael and Juty, Nick and Juvan, Peter and Koch, Maximilian and Laibe, Camille and Li, Jing-Woei and Licata, Luana and Mareuil, Fabien and Mičetić, Ivan and Friborg, Rune Møllegaard and Moretti, Sebastien and Morris, Chris and Möller, Steffen and Nenadic, Aleksandra and Peterson, Hedi and Profiti, Giuseppe and Rice, Peter and Romano, Paolo and Roncaglia, Paola and Saidi, Rabie and Schafferhans, Andrea and Schwämmle, Veit and Smith, Callum and Sperotto, Maria Maddalena and Stockinger, Heinz and Vařeková, Radka Svobodová and Tosatto, Silvio C.E. and de la Torre, Victor and Uva, Paolo and Via, Allegra and Yachdav, Guy and Zambelli, Federico and Vriend, Gert and Rost, Burkhard and Parkinson, Helen and Løngreen, Peter and Brunak, Søren},
	month = jan,
	year = {2016},
	pages = {D38--D47},
}

@article{WELDER20181130,
    title = {Spatio-temporal optimization of a future energy system for power-to-hydrogen applications in Germany},
    journal = {Energy},
    volume = {158},
    pages = {1130-1149},
    year = {2018},
    issn = {0360-5442},
    doi = {https://doi.org/10.1016/j.energy.2018.05.059},
    url = {https://www.sciencedirect.com/science/article/pii/S036054421830879X},
    author = {Lara Welder and D.Severin Ryberg and Leander Kotzur and Thomas Grube and Martin Robinius and Detlef Stolten}
}

@article{HOFFMANN2021117825,
    title = {Typical periods or typical time steps? A multi-model analysis to determine the optimal temporal aggregation for energy system models},
    journal = {Applied Energy},
    volume = {304},
    pages = {117825},
    year = {2021},
    issn = {0306-2619},
    doi = {https://doi.org/10.1016/j.apenergy.2021.117825},
    url = {https://www.sciencedirect.com/science/article/pii/S0306261921011545},
    author = {Maximilian Hoffmann and Jan Priesmann and Lars Nolting and Aaron Praktiknjo and Leander Kotzur and Detlef Stolten}
}

@article{leeTenSimpleRules2018,
    title = {Ten Simple Rules for Documenting Scientific Software},
    author = {Lee, Benjamin D.},
    date = {2018-12-20},
    journaltitle = {PLOS Computational Biology},
    shortjournal = {PLOS Computational Biology},
    volume = {14},
    number = {12},
    pages = {e1006561},
    publisher = {Public Library of Science},
    issn = {1553-7358},
    doi = {10.1371/journal.pcbi.1006561},
    url = {https://journals.plos.org/ploscompbiol/article?id=10.1371/journal.pcbi.1006561},
    urldate = {2022-11-18},
    langid = {english}
}

@book{zeng_metadata_2022,
	address = {London, Great Britain},
	edition = {Third edition},
	title = {Metadata},
	isbn = {978-1-78330-588-9},
	author = {Zeng, Marcia Lei and Qin, Jian},
	publisher = {Facet Publishing},
	year = {2022}
}

@article{barker2022introducing,
    title={Introducing the FAIR Principles for research software},
    author={Barker, Michelle and Chue Hong, Neil P and Katz, Daniel S and Lamprecht, Anna-Lena and Martinez-Ortiz, Carlos and Psomopoulos, Fotis and Harrow, Jennifer and Castro, Leyla Jael and Gruenpeter, Morane and Martinez, Paula Andrea and others},
    journal={Scientific Data},
    volume={9},
    number={1},
    pages={1--6},
    year={2022},
    publisher={Nature Publishing Group}
}

@article{smith2016software,
    title={Software citation principles},
    author={Smith, Arfon M and Katz, Daniel S and Niemeyer, Kyle E},
    journal={PeerJ Computer Science},
    volume={2},
    pages={e86},
    year={2016},
    publisher={PeerJ Inc.}
}

@article{anzt2021crediting,
    title={Crediting pull requests to open source research software as an academic contribution},
    author={Anzt, Hartwig and Kuehn, Eileen and Flegar, Goran},
    journal={Journal of Computational Science},
    volume={49},
    pages={101278},
    year={2021},
    publisher={Elsevier}
}

@misc{IAMCsPyamDataModel,
	title = {IAMC's Pyam Data Model},
	url = {https://pyam-iamc.readthedocs.io/en/stable/data.html},
	urldate = {2023-01-09},
}

@misc{OWL-S,
	title = {OWL-S: Semantic Markup for Web Services},
	url = {https://www.w3.org/Submission/OWL-S/},
	urldate = {2023-01-16},
}

@misc{WSDL,
	title = {Web Services Description Language (WSDL) Version 2.0 Part 1: Core Language - W3C Recommendation 26 June 2007},
	url = {https://www.w3.org/TR/wsdl/},
	urldate = {2023-01-18},
}

@misc{WADL,
	title = {Web Application Description Language - W3C Member Submission 31 August 2009},
	url = {https://www.w3.org/Submission/wadl/},
	urldate = {2023-01-18},
}

@misc{OpenAPI,
	title = {OpenAPI Specification v3.1.0},
	url = {https://spec.openapis.org/oas/v3.1.0},
	urldate = {2023-01-18},
}

@misc{FMI,
	title = {Functional Mock-up Interface Specification, version 3.0, 2022-05-10},
	url = {https://fmi-standard.org/docs/3.0/},
	urldate = {2023-01-18},
}

@misc{DublinCore,
	title = {Dublin Core Metadata Initiative (DCMI) Metadata Terms},
	url = {https://www.dublincore.org/specifications/dublin-core/dcmi-terms/},
	urldate = {2022-08-11},
}

@misc{DataCite,
	title = {DataCite Metadata Schema},
	url = {https://schema.datacite.org/},
	urldate = {2022-08-11},
}

@misc{CodeMeta,
	title = {The CodeMeta Project},
	url = {https://codemeta.github.io/},
	urldate = {2023-01-16},
}

@misc{CodeMetaGenerator,
	title = {CodeMeta generator},
	url = {https://codemeta.github.io/codemeta-generator/},
	urldate = {2023-01-27},
}

@misc{SoftwareDescriptionOntology,
	title = {The Software Description Ontology - Release May 3rd, 2021},
	url = {https://w3id.org/okn/o/sd},
	urldate = {2023-01-16},
}

@misc{OntoSoftPortal,
	title = {OntoSoft Portal},
	url = {https://www.ontosoft.org/portal/},
	urldate = {2023-01-09},
}

@misc{MINT,
	title = {MINT Model Explorer},
	url = {http://models.mint.isi.edu},
	urldate = {2023-01-16},
}

@misc{BioTools,
	title = {bio.tools},
	url = {http://bio.tools},
	urldate = {2023-01-16},
}

@misc{FINEDocumentation,
	title = {Welcome to FINE's documentation! FINE - A Framework for Integrated Energy System Assessment},
	url = {https://vsa-fine.readthedocs.io/en/master/},
	urldate = {2023-01-20},
}

@misc{PyPSADocumentation,
	title = {PyPSA: Python for Power System Analysis},
	url = {https://pypsa.readthedocs.io/en/latest/index.html},
	urldate = {2023-05-26},
}

@misc{FINEsEnergySystemModel,
	title = {FINE's Energy System Model Class},
	url = {https://vsa-fine.readthedocs.io/en/master/sourceCodeDocumentation/energySystemModelDoc.html},
	urldate = {2023-01-09},
}

@misc{openmodWikiOpenModels,
	title = {openmod initiative's Wiki - Open Models},
	url = {https://wiki.openmod-initiative.org/wiki/Open_Models},
	urldate = {2023-01-16},
}

@misc{OEPModelFactsheets,
	title = {Open Energy Platform (OEP) - Model Factsheets},
	url = {https://openenergy-platform.org/factsheets/models/},
	urldate = {2023-01-16},
}

@misc{FINEOEPModelFactsheet,
	title = {OEP Framework Factsheet - Framework Framework for Integrated Energy System Assessment (FINE)},
	url = {https://openenergy-platform.org/factsheets/frameworks/164/},
	urldate = {2023-02-18},
}

@misc{OEKG,
	title = {Open Energy Knowledge Graph (OEKG)},
	url = {https://github.com/OpenEnergyPlatform/oekg},
	urldate = {2023-01-30},
}

@misc{Javadoc,
	title = {Javadoc Tool},
	url = {https://www.oracle.com/java/technologies/javase/javadoc.html},
	urldate = {2023-01-23},
}

@misc{Perldoc,
	title = {Perldoc Browser 5.36.0},
	url = {https://perldoc.perl.org/perlpod},
	urldate = {2023-01-23},
}

@misc{Doxygen,
	title = {Doxygen},
	url = {https://www.doxygen.nl/},
	urldate = {2023-01-23},
}

@misc{SphinxNapoleon,
	title = {sphinx.ext.napoleon – Support for NumPy and Google style docstrings},
	url = {https://www.sphinx-doc.org/en/master/usage/extensions/napoleon.html},
	urldate = {2023-01-23},
}

@misc{MkDocs,
	title = {MkDocs - Project documentation with Markdown.},
	url = {https://www.mkdocs.org/},
	urldate = {2023-01-23},
}

@misc{Sphinx,
	title = {Sphinx - Python Documentation Generator},
	url = {https://www.sphinx-doc.org/en/master/},
	urldate = {2023-01-23},
}

@misc{Roxygen2,
	title = {roxygen2 7.2.3},
	url = {https://roxygen2.r-lib.org/},
	urldate = {2023-01-23},
}

@misc{Swagger,
	title = {Swagger - API Documentation and Design Tools for Teams},
	url = {https://swagger.io/},
	urldate = {2023-01-23},
}

@misc{GitHub,
	title = {GitHub - Let's build from here},
	url = {https://github.com/},
	urldate = {2023-01-24},
}

@misc{GitLab,
	title = {GitLab - The DevSecOps Platform},
	url = {https://gitlab.com/},
	urldate = {2023-01-24},
}

@misc{GithubPages,
	title = {Github Pages - Websites for you and your projects},
	url = {https://pages.github.com/},
	urldate = {2023-01-25},
}

@misc{NPM,
	title = {NPM - Node Package Manager},
	url = {https://www.npmjs.com/},
	urldate = {2023-01-25},
}

@misc{Sourceforge,
	title = {Sourceforge - The Complete Open-Source and Business Software Platform},
	url = {https://sourceforge.net/},
	urldate = {2023-01-25},
}

@misc{Bitbucket,
	title = {Bitbucket - Code and CI/CD, built for teams using Jira},
	url = {https://bitbucket.org/},
	urldate = {2023-01-24},
}

@misc{Anaconda,
	title = {Package repository for anaconda},
	url = {https://anaconda.org/anaconda/repo},
	urldate = {2023-01-24},
}

@misc{Maven,
	title = {Maven Central Repository Search},
	url = {https://search.maven.org/},
	urldate = {2023-01-24},
}

@misc{SwaggerHub,
	title = {SwaggerHub - Search public APIs and Domains in SwaggerHub},
	url = {https://app.swaggerhub.com/search},
	urldate = {2023-01-30},
}

@misc{ReadTheDocs,
	title = {Read the Docs - Dokumentation erstellen, hosten und durchsuchen.},
	url = {https://readthedocs.org/},
	urldate = {2023-01-24},
}

@misc{GitBook,
	title = {GitBook - Where technical teams document.},
	url = {https://www.gitbook.com/},
	urldate = {2023-01-24},
}

@misc{CRAN,
	title = {The Comprehensive R Archive Network (CRAN)},
	url = {https://cran.r-project.org/},
	urldate = {2023-01-24},
}

@misc{ORKG,
	title = {Open Research Knowledge Graph (ORKG)},
	url = {https://orkg.org/},
	urldate = {2023-01-27},
}

@misc{PyPi,
	title = {PyPi - Der Python Package Index},
	url = {https://pypi.org/},
	urldate = {2023-01-26},
}

@misc{OpenAPITools,
	title = {OpenAPI.Tools},
	url = {https://openapi.tools/},
	urldate = {2023-01-27},
}

@misc{DataDesc,
	title = {DataDesc - An ecosystem for machine-actionable software metadata},
	url = {https://github.com/FZJ-IEK3-VSA/DataDesc},
	urldate = {2023-01-31},
}

@article{booshehri2021introducing,
    title={Introducing the Open Energy Ontology: Enhancing data interpretation and interfacing in energy systems analysis},
    author={Booshehri, Meisam and Emele, Lukas and Fl{\"u}gel, Simon and F{\"o}rster, Hannah and Frey, Johannes and Frey, Ulrich and Glauer, Martin and Hastings, Janna and Hofmann, Christian and Hoyer-Klick, Carsten and others},
    journal={Energy and AI},
    volume={5},
    pages={100074},
    year={2021},
    publisher={Elsevier}
}

@article{heiler1995semantic,
    title={Semantic interoperability},
    author={Heiler, Sandra},
    journal={ACM Computing Surveys (CSUR)},
    volume={27},
    number={2},
    pages={271--273},
    year={1995},
    publisher={ACM New York, NY, USA}
}

@misc{FINERepository,
	title = {FINE - Framework for Integrated Energy System Assessment},
	url = {https://github.com/FZJ-IEK3-VSA/FINE},
	urldate = {2023-02-11},
}

@misc{tsamRepository,
	title = {tsam - Time Series Aggregation Module},
	url = {https://github.com/FZJ-IEK3-VSA/tsam},
	urldate = {2023-02-21},
}

@article{kelley2021framework,
  title={A framework for creating knowledge graphs of scientific software metadata},
  author={Kelley, Aidan and Garijo, Daniel},
  journal={Quantitative Science Studies},
  volume={2},
  number={4},
  pages={1423--1446},
  year={2021},
  publisher={MIT Press One Rogers Street, Cambridge, MA 02142-1209, USA journals-info~…}
}

@article{habermann2020metadata,
  title={Metadata and reuse: Antidotes to information entropy},
  author={Habermann, Ted},
  journal={Patterns},
  volume={1},
  number={1},
  pages={100004},
  year={2020},
  publisher={Elsevier}
}

@article{druskat2022software,
  title={Software publications with rich metadata: state of the art, automated workflows and HERMES concept},
  author={Druskat, Stephan and Bertuch, Oliver and Juckeland, Guido and Knodel, Oliver and Schlauch, Tobias},
  journal={arXiv preprint arXiv:2201.09015},
  year={2022}
}

\end{document}